\documentstyle [12pt]{article}
\begin{document}

\setcounter{page}{1}
\title{The RNO framework to the $\sigma$-particle mass}
\author{G.A.Kozlov\\
{\em Bogoliubov Laboratory of Theoretical Physics}\\
\em Joint Institute for Nuclear Research\\
\em 141980 Dubna, Moscow Region, Russia}
\date{}
\maketitle
\begin{abstract}

{\small The "standard" iso-singlet scalar particle $\sigma$ is
reconsidered in the reduced normal-ordering (RNO) framework to the
effective SU(2) theory.  Recent reanalysis of the $\pi\pi$-phase
 shift [1] is used.\\ }

\end{abstract}
1. There has been a long-standing history of the existence of chiral
partner of $\pi$-meson, $\sigma$-meson [2], or against its existence
[3].  The $\sigma$-particle is an iso-singlet scalar one, and it is
required to exist in the effective Lagrangian density in the simple
linear representation [4] of the chiral symmetry. There is no clear
evidence, however, of the existence of a scalar resonance with mass
around 0.5 GeV. The importance of the role of the $\sigma$-meson in
nuclear physics (nuclear forces with one-boson exchange potential
[5]), regarding the point of view of spontaneous breaking of chiral
symmetry is well known. For example, a phase transition of the QCD
vacuum is considered as a key to explain the chiral symmetry
breaking. Based on the Nambu-Jona-Lasinio model, Hatsuda and Kunihiro
[6] presented the pion in their QCD-motivated effective theory as a
phase fluctuation of an order parameter $<\bar{q}q>$, the thermal
average of the vacuum expectation value [7]. The $\sigma$-meson is an
amplitude fluctuation of this order parameter, and it plays a similar
role as the Higgs remnant in the Nambu-Goldstone phase [7].

   The analysis of $\pi\pi$ phase-shift obtained in a
CERN-M$\ddot{u}$nich experiment [8] leads to the point of view of
nonexistence of $\sigma$ as a resonant particle. Meanwhile, the
fundamental importance of $\sigma$ in relation with the dynamical
breaking of chiral symmetry is discussed extensively [9]. A new more
rigorous analysis of $\pi\pi$ phase-shift has been quite necessary.
The Japanese group [1,10] performed this by applying an original
method of analysis, the interfering-amplitude (IA) method. As a main
result, an evidence for the existence of $\sigma$-meson has been
shown from the characteristic behavior of the
$\pi\pi\rightarrow\pi\pi$ phase shift under the $K\bar{K}$
threshold.\\
2. In this short paper we shall reconsider the SU(2) linear
$\sigma$-model [4,11] in an ordinary perturbation expansion, taking
the RNO interaction Lagrangian density (LD). We start with the
following LD written in the standard form:
\begin{eqnarray}
\label{e1}
L=\frac{1}{2}\partial_{\mu}\phi\partial^{\mu}\phi+
\frac{1}{2}\partial_{\mu}
\sigma_{0}\partial^{\mu}\sigma_{0}-\frac{1}{2}\mu_{0}^{2}(\phi^2+
\sigma_{0}^2)-\frac{\lambda}
{4}{(\phi^2+\sigma_{0}^2)}^2+f_{\pi}\,m_{\pi}^2\,\sigma_{0}\ ,
\end{eqnarray}
where $\phi(x)$ represents the iso-triplet $\pi$-field, while
$\sigma_{0}(x)$
stands for the field of $\sigma$-particle; $f_{\pi}$ and $m_{\pi}$
are the decay constant and mass of the pion, respectively. To treat
the case of spontaneous symmetry breaking, one should rewrite the
$\sigma$-meson field as $\sigma_{0}(x)=\sigma(x)+v$, where $v\equiv
<\sigma_{0}(x)>_{0}\not= 0$.  We divide the total LD (~\ref{e1}~) as
$$L=L_{0}+L_{I}+L_{\pi\sigma}\, $$
where
$$L_{0}=\frac{1}{2}\partial_{\mu}\phi\partial^{\mu}\phi+\frac{1}{2}
\partial_{\mu}
\sigma\partial^{\mu}\sigma-\frac{1}{2}\mu^{2}\phi^2-
\frac{1}{2}m^2\sigma^2
-\lambda(v\cdot\sigma)^2\ , $$

\begin{eqnarray}
\label{e2}
L_{I}=-\frac{\lambda}{4}(\phi^2)^2-\frac{\lambda}{4}(\sigma^2)^2-\lambda(v
\cdot\sigma)(\sigma^2+\phi^2)+
+\frac{1}{2}\phi^2(\mu^2-\mu_{0}^2-\lambda\,v^2)+ \cr
+\frac{1}{2}[m^2-\mu_{0}^2-\lambda(\phi^2+v^2)]\sigma^2-
(\mu_{0}^2+\lambda\,v^2)(v\cdot\sigma)-\frac{1}{2}\mu_{0}^2\,v^2-\frac
{\lambda}{4}(v^2)^2\ ,
\end{eqnarray}
$$L_{\pi\sigma}=f_{\pi}\,m_{\pi}^2\,(\sigma+v)\ .$$
Here, new parameters $m$ and $\mu$ are both arbitrary. Our model is
quite suitable and well fits if we rewrite (~\ref{e2}~) in terms of
the reduced form of normal-ordered field operators. In the standard
perturbation expansion in the framework of the interaction
representation, one has to be precise in taking a normal-ordered
interaction LD :$L_{I}$: instead of $L_{I}$. Following Nakano [12] we
take $L_{I}$ but not :$L_{I}$: although one presents $L_{I}$
(~\ref{e2}~) in terms of certain normal-ordered operators. It has
been noted in [12] that the partial normal ordering procedure has the
advantage because it is needless to take into account tadpole
diagrams.

   The trial point in the method given here is that only field
quadratic factors are defined in terms of the normal-ordered
operators of the $\sigma$-field. One of the essential steps in our
approach is that we rewrite the quadratic field factor $\sigma^2(x)$
as
\begin{eqnarray}
\label{e3}
\sigma^2(x)=:\sigma^2(x):+\Delta(m,x)\ ,
\end{eqnarray}
where the two-point Wightman function is [13]
\begin{eqnarray}
\label{e4}
\Delta(m,x)=\frac{-i}{4\,\pi}\epsilon (x^0)\,\delta (x^2)
+\frac{m}{8\,\pi\,
\sqrt{x^2}}\,\Theta(x^2)\cdot \cr
\cdot \left [N_{1}(m\,\sqrt{x^2})+i\,\epsilon(x^0)\,J_{1}(m\,
\sqrt{x^2})
\right ]+\frac{m}{4\,\pi^2\,\sqrt{-x^2}}\,\Theta(-x^2)\,
K_{1}(m\sqrt{-x^2})\ ,
\end{eqnarray}
$N_{1}(z), J_{1}(z)$ and $K_{1}(z)$ are the Neumann functions, Bessel
functions and the Bessel functions of imaginary argument,
respectively. We are not going to rewrite the quartic field factor
$(\sigma^2)^2$ in (~\ref{e2}~) in terms of :$(\sigma^2)^2$:. Hence,
we do only the quadratic normal-ordered procedure. In fact, our
estimation on the magnitudes is somewhat formal since the
$\Delta(m,x)$-function (~\ref{e4}~) contains the divergence coming
from short distances and the logarithmic divergences because of too
light $\sigma$ as
$$ N_{1}(m\,x)=\frac{2}{\pi}\,J_{1}(m\,x)\cdot\ln(m\,x/2)
\,\,\,\, x^2>0 \ .$$
Substituting (~\ref{e3}~) into (~\ref{e2}~) we obtain the expression
of $L_{I}$ on which we develop perturbative expansion
$$L_{I}=-\frac{\lambda}{4}(:\sigma^{2}:)^2-
\lambda(v\cdot\sigma):\sigma^2:+
\frac{1}{2}\,\delta\mu_{0}^2:\sigma^2:\,-$$
$$-[\mu_{0}^2+\lambda(v^2+\Delta(m,x)+\phi^2)](v\cdot\sigma)
-\frac{1}{2}(
\mu_{0}^2+\lambda\phi^2)v^2\,-$$
$$-\frac{\lambda}{4}[(v^2)^2+\Delta^2(m,x)]+\frac{1}{2}[m^2-\mu_{0}^2-
\lambda(v^2+\phi^2)]\Delta(m,x)\,-$$
$$-\frac{\lambda}{4}(\phi^2)^2+\frac{1}{2}\phi^2(\mu^2-\mu_{0}^2)
-\frac{1}{2}
\lambda:\sigma^2:\phi^2\ ,$$
where we denote $$\delta\mu_{0}^2\equiv
m^2-\mu_{0}^2-\lambda[v^2+\Delta(m,x)]\ .$$ Note that $v$ stands for
the vacuum expectation value (v.e.v.) of the $\sigma$-field
determined from the condition that terms linear in the real physical
$\sigma$-fields should disappear from the initial LD (~\ref{e1}~) as
\begin{eqnarray}
\label{e5}
\{\mu_{0}^2+\lambda[v^2+\Delta(m,x)]\}\,v\,-\,f_{\pi}\,m_{\pi}^2=0\ .
\end{eqnarray}
Since the model has to be independent of $m^2$, one can fix this fact
such that $\delta\mu_{0}^2$\,=\,0, i.e. $m=0$, taking into account
(~\ref{e5}~) in the chiral limit $m_{\pi}$=0 ($v\not=$0).

  At the present stage of perturbation, the two-point function does not
contain loop corrections, and the physical masses of $\sigma$ could be
extracted using the unperturbed LD
$$L_{0}=\frac{1}{2}\partial_{\mu}\phi\partial^{\mu}\phi+\frac{1}{2}
\partial_{\mu}
\sigma\partial^{\mu}\sigma-\frac{1}{2}\mu^{2}\phi^2
-\frac{1}{2}m^2\sigma^a
\left(\delta_{ab}-\frac{v_{a}\,v_{b}}{v^2}\right)\sigma^{b}\,-$$
$$-\frac{1}{2}(m^2+2\,\lambda\,v^2)\,\sigma^a\,\frac{v_{a}
\,v_{b}}{v^2}\,
\sigma^b\ .$$
If $\sigma(x)$ is a scalar multiplet of O(N), one can immediately
find (N-1) Goldstone-like particles and one massive boson with the
following corresponding masses:
$$m_{1}^2\,=\,m^2\,=\,0 $$
and
$$m_{2}^2\,=\,2\,\lambda\,v^2\ ,$$
where the zeroth mass is a direct consequence of the Goldstone
theorem.

 3. Let us rewrite LD (~\ref{e1}~) in terms of the normal-ordered
$:\sigma^2:$-operators
$$L\simeq\frac{1}{2}\partial_{\mu}\phi\partial^{\mu}\phi+\frac{1}{2}
\partial_{\mu}
\sigma\partial^{\mu}\sigma-\frac{1}{2}m_{\pi}^{2}\,\phi^2-
\frac{1}{2}m_{\sigma}^2:\sigma^2:-$$
$$-\lambda(v\cdot\sigma)(:\sigma^2:+\phi^2)-\frac{\lambda}{4}
(:\sigma^2:+
\phi^2)^2-\frac{\lambda}{4}(v^2)^2-\frac{1}{2}\mu_{0}^2\,v^2
+f_{\pi}\,m_{\pi}
^2\,v\ ,$$
where
$$m_{\sigma}^2\,=\,m_{\pi}^2\,+\,2\,\lambda\,v^2\ ,$$
\begin{eqnarray}
\label{e6}
m_{\pi}^2\,=\,\mu_{0}^2\,+\,\lambda[v^2\,+\,\Delta(m,x)]\ .
\end{eqnarray}
Here, we have the fact used, that all the singularities in the
$\Delta(m,x)$- function (~\ref{e4}~) are lying on the light front at
$x^2\,=\,0$, while this function $\Delta(m,x)$ behaves like ${\vert
x^2\vert}^{-3/4}\,\exp(- m\,\vert x\vert )$ and $ {\vert x\vert
}^{-3/2}$ at the space- and time-like infinity, respectively [13].
Comparing (~\ref{e6}~) with the condition (~\ref{e5}~) we can
formally find, that v.e.v. $v$ is just equal to $f_{\pi}$, and thus,
\begin{eqnarray}
\label{e7}
m_{\sigma}^2\,=\,m_{\pi}^2\,+\,2\,\lambda\,f_{\pi}^2\ .
\end {eqnarray}
It has been remarked in [10] that the result of one-loop calculation
of the linear $\sigma$-model [14,15] gives corrections to the tree
approximation, which are small ones. For numerical estimation of the
 $\sigma$-mass at the tree level, let us rewrite (~\ref{e7}~) in the
following form:
\begin{eqnarray}
\label{e8}
m_{\sigma}\,=\,{(m_{\pi}^2\,+\,2\,f_{\pi}\,g_{\pi\sigma})}^{1/2}\ ,
\end{eqnarray}
where the $\pi\sigma$-coupling constant $g_{\pi\sigma}\,
=\,\lambda\,v$ is
chosen as result of the reanalysis of the $\pi\pi$ phase-shift
analysis based on the IA-method [1,10].
Taking the central value in the estimation of the relevant parameter
$g_{\pi\sigma}\,=\,1.362\pm0.005~GeV$, given in [10], we find for
$m_{\sigma}$
$$m_{\sigma}\,=\,522~MeV\ ,$$
which is to be compared with other theoretical estimations [2] and
the relevant parameter $m_{\sigma}=553.3\pm 0.5~MeV$ obtained in
[10]. Let us check the relation (~\ref{e8}~) with the value
$g_{\pi\sigma}=1.54~GeV$ [10]. The result is $m_{\sigma}=552.7~MeV$.
In the following paper, we shall examine $m_{\sigma}$ using the
special class of an extended low-dimensional scalar effective theory
containing an additional light scalar described by the dipole-type
field obeying the higher-order equation [16].

\end{document}